\newif\ifAlpha \Alphafalse

\newif \ifhyperlinks    \hyperlinkstrue

\newif \ifDraft         \Drafttrue
\newif \ifAnon	\Anontrue
\newif \ifFinal		\Finalfalse

\ifFinal\Anonfalse\fi
\Draftfalse
\Anonfalse

  \documentclass[letterpaper,10pt,twocolumn]{article}
\pdfoutput=1
\usepackage[pdftex]{graphicx}
  \usepackage{usenix}

  \newif \ifWebversion    \Webversiontrue 
  \usepackage[authoryear]{natbib}
\usepackage{paralist}           
\usepackage{verbatim}           
\usepackage[dvipsnames]{xcolor}
\usepackage{colortbl}
\usepackage{subfig}
\newcommand{\autorefsub}[2]{\autoref{#1}\subref{#2}}

\usepackage{xspace}             
\newcommand{\composite}{\textrm{Composite}\xspace}

\usepackage{balance}            

\usepackage[pdftex]{graphicx}
\setkeys{Gin}{keepaspectratio=true,clip=true,draft=false,width=\linewidth}
\graphicspath{{./imgs/}}

\ifDraft
  \usepackage{draftwatermark}
  \SetWatermarkLightness{0.85}
  \newcommand{\Comment}[1]{\textbf{\textsl{#1}}}
  \newenvironment{LongComment}[1] 
    {\begingroup\par\noindent\slshape \textbf{Begin Comment[#1]}\par}
    {\par\noindent\textbf{End Comment}\endgroup\par}
  \newcommand{\FIXME}[1]{\textbf{\textsl{FIXME: #1}}}
  \newcommand{\TODO}[1]{\textbf{\textsl{TODO: #1}}}
\else
  \newcommand{\Comment}[1]{\relax}
  
  \newcommand{\FIXME}[1]{\relax}
  \newcommand{\TODO}[1]{\relax}
\fi

\newcommand{\bbb}[1]{\Comment{#1 [Bj\"{o}rn]}}

\newcommand{\code}[1]{\texttt{#1}\xspace}
\newcommand{\crit}[1]{\textsc{#1}\xspace}
\usepackage{multirow}

\usepackage[pdftex]{hyperref}
\ifhyperlinks\else   
  \hypersetup{nolinks=true}
\fi

\begin{document}
  \sloppy

  \renewcommand{\sectionautorefname}{Section}
  \renewcommand{\subsectionautorefname}{Section}
  \renewcommand{\subsubsectionautorefname}{Section}
  \renewcommand{\appendixautorefname}{Appendix}
  \renewcommand{\Hfootnoteautorefname}{Footnote}
  \newcommand{\Htextbf}[1]{\textbf{\hyperpage{#1}}}

  \title{It's Time: OS Mechanisms for Enforcing Asymmetric Temporal
    Integrity}
  \author{
    \ifAnon 
      Paper \#180, \pageref{p:last} pages of 12 allowed
    \else
      Anna Lyons, Gernot Heiser\\
      DATA61 (formerly NICTA) and UNSW Australia\\
      \href{mailto:anna.lyons@data61.csiro.au}{\{anna.lyons,gernot.heiser\}@data61.csiro.au}
    \fi
  }
  \maketitle

\ifFinal
  \pagestyle{empty}
\fi




  \subsection*{Abstract}
  Mixed-criticality systems combine real-time components of different
  levels of criticality, i.e.\ severity of failure, on the same
  processor, in order to obtain good resource utilisation. They must
  guarantee deadlines of highly-critical tasks at the expense of
  lower-criticality ones in the case of overload. Present operating
  systems provide inadequate support for this kind of system, which is
  of growing importance in avionics and other verticals. We
  present an approach that provides the required asymmetric integrity
  and its implementation in the high-assurance seL4 microkernel.

\section{Introduction}\label{s:intro}

Traditionally, critical real-time systems use dedicated
microcontrollers for each function. With increasing functionality and
complexity of cyber-physical and other real-time systems, this is
creating space, weight and power (SWaP) problems, which force
consolidation onto a smaller number of more powerful processors. For
example, top-end cars reached 100 processors a few years ago
\citep{Hergenhan_Heiser_08}; with the robust packaging and wiring required for
vehicle electronics, the SWaP problem is obvious, and a driver for the
adoption of multitasking OSes \citep{AUTOSAR:AUTOSAR_Spec_42}.

The potential for consolidation is limited unless it is possible
to safely co-host functions of different \emph{criticality}, where
criticality is a well-established notion that represents the severity
of failure \citep{DO178B}. Certification standards require that safe operation of a
particular component must not depend on any less-critical components
\citep{ARINC653}.

Such \emph{mixed-criticality systems} (MCS) are becoming the norm in
avionics, but presently in a very restricted form: the system is
orthogonally portioned spatially and temporally, and partitions are
scheduled round-robin with fixed time slices \citep{ARINC653}. This
limits integration and cross-partition communication, and implies long
interrupt latencies and poor resource utilisation. The simple
partitioning approach will not meet the requirements of future
mixed-criticality systems \citep{Barhorst_BBHPSSSSU_09}.

Fundamental to good resource utilisation in MCS is the ability to
over-commit safely: The system's core \emph{integrity property} is
that deadlines of the highest criticality tasks must be guaranteed,
meaning that there is always time to let such tasks execute their full \emph{worst-case
execution time} (WCET). This may be orders of magnitude larger than the
typical execution time, and computation of safe WCET bounds for
non-trivial software tends to be highly pessimistic \citep{Wilhelm_EEHTWBFHMMPPSS_08}. This
means that most of the time the highly-critical components leave plenty of slack, which
should be available to less critical components, but must be available
to the critical component when needed.

\begin{figure}[t]
  \centering
  \includegraphics[width=0.8\columnwidth]{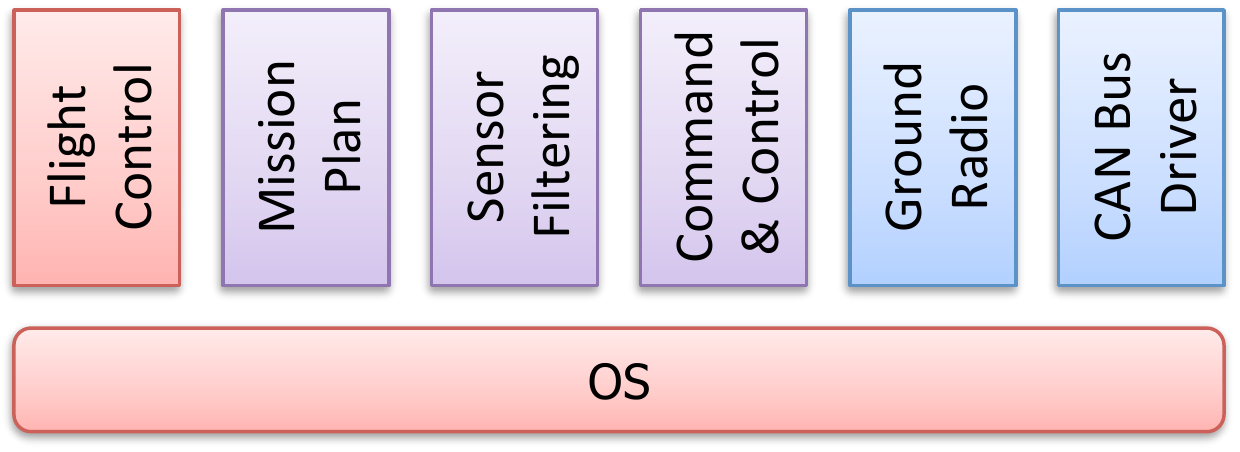}
  \caption{Highly simplified autonomous aerial vehicle architecture red is most critical,
    blue least.}
  \label{f:aav}
\end{figure}

Such a system needs support for downgrading timeliness guarantees selectively,
least critical ones first. In general, this cannot be achieved by
simply giving the most critical tasks the highest priority. Consider
the simplified architecture of an \emph{autonomous aerial vehicle} (AAV)
in \autoref{f:aav}. The most critical component is the low-level
flight control, which keeps the vehicle stable and moving towards a
waypoint. It executes every 100\,ms and normally takes about 10\,ms but
has a WCET of 70\,ms. Next critical are the mission plan, sensor
filtering and C\&C components, which have execution rates of between 1
and 10\,Hz, normally run for a combined 200\,ms every second but have a combined
WCET of 500\,ms per second. The CAN bus, which connects a video camera
and various
sensors of secondary importance, can deliver packets every 12.5\,\(\mu\)s
and does not buffer.

If the critical components are given higher
priority than the CAN driver, it will drop many
packets even during normal operation, despite the system
having sufficient headroom to run everything.
The standard realtime (RT) scheduling approach is \emph{rate-monotonic
priority assignment} (RMPA) \citep{Liu_Layland_73}, which gives
highest priority to tasks with the shortest periods. RMPA is easy to
analyse and known to be optimal for fixed priorities; it is highly desirable to retain it
for MCS.

A further complication is that components of different criticality
must be able to communicate, and access shared data
\citep{Burns_Baruah_13}. For example, the
AAV's mission plan defines the waypoints to be used by the flight
control, including some fail-safe return-home path in case the AAV
loses ground-station connectivity. It is updated by the ground station via the command and
control (C\&C) component,
and amended by the sensor filtering component for obstacle avoidance;
the latter component receives input from various sensors, including
camera and other sensor input via the CAN bus.
Such communication, including concurrency control between
components accessing the same data, must be possible while
guaranteeing critical deadlines.

In summary, an OS for mixed-criticality systems must:
\begin{compactitem}
\item provide high-assurance spatial and temporal isolation, to allow
  critical components to be assured independently of less critical
  ones;
\item decouple criticality from priority, to ensure critical,
  low-rate threads meet their deadlines;
\item provide mechanisms that allow analysing the timeliness of
  critical tasks, even if they communicate with less critical ones;
\item have well-understood temporal behaviour, especially
  bounded and known WCET for all operations;
\item be highly assured for correct operation.
\end{compactitem}

No such OS exists to date.
We present the design and implementation of such an OS, based on the
seL4 microkernel for single-core systems. seL4 is an attractive starting point, as it is a
high-assurance OS kernel that has been 
comprehensively verified \citep{Klein_AEMSKH_14}, and is the first and
still only protected-mode OS in the literature with a complete and sound WCET analysis
\citep{Blackham_SCRH_11}. 

We do \emph{not} claim to have invented new scheduling models or
theory. In fact, the system we present in \autoref{s:design} is, as
scheduling theory goes, known as \emph{static mixed criticality}
\citep{Baruah_BD_11}. Our claims are about practical systems, specifically:
\begin{compactenum}
\item the design of a low-overhead temporal resource management model
  that is based on a small number of simple, policy-free mechanisms,
  suitable for a high-assurance implementation,
  matches the above requirements of MCS but also supports a wide
  range of other uses (\autoref{s:design});
\item its implementation in the seL4 microkernel in a way that retains
  seL4's general-purpose
  nature and verifiability\footnote{We have not
    formally re-verified the modified kernel, and only claim that our
    modifications are moderate in terms of kernel changes and no more
    difficult to verify than the baseline kernel.} (\autoref{s:impl});
\item an evaluation that demonstrates that the modifications do not
  unduly impact seL4's performance, and support low-overhead
  implementations of different real-time and best-effort scheduling
  models (\autoref{s:eval}).
\end{compactenum}

\section{Background and Related Work\label{s:background}}

In the rest of this paper, and this section specifically, we talk
about general real-time concepts as well as OS
abstraction. Specifically there are two related concepts relating to
the execution model. We will use the term \emph{task} in the sense
established in the RT community, namely \emph{a set of related jobs
  which jointly provide some system function}, where a job is \emph{a
  unit of work that is scheduled and executed by the system}
\citep{Liu:rts}. We use the term (kernel-scheduled) \emph{thread} to refer to the
execution abstraction familiar to the OS community.  The term ``job'',
which we will not use further, corresponds to a unit of work that is conducted
by a thread, while ``task'' maps onto a thread, plus code and data.

In short, we will use ``task'' when referring to general RT issues,
and ``thread'' when talking about a specific OS concept. In practice,
the terms are largely interchangeable.

\subsection{Scheduling models vs.\ mixed criticality}\label{s:mc-sched}

RT scheduling generally assumes \emph{periodic tasks}, which maps well
onto typical control systems, where different activities execute
periodically albeit with different periods. Non-periodic
(``sporadic'', i.e.\ interrupt-driven) tasks are incorporated in such a model by requiring
a defined \emph{minimum arrival time}, corresponding to a maximum
interrupt rate, which is used as the task's period for the
schedulability analysis. RT tasks have a \emph{deadline} by which a
computation must be finished. The general assumption is that deadlines
are implicit, meaning the deadline is the end of the period.

As discussed in the introduction, the ability to overload, while
guaranteeing critical deadlines, is core to the notion of
MCS. Classical RT scheduling approaches have a notion of (fixed or
dynamic) priority as the sole determinant of access to CPU time, with
equal-priority tasks (if permitted) being (preemptively or non-preemptively)
scheduled FIFO. If the system is overloaded, this means that
the lowest-priority deadlines are missed. When using RMPA, this
victimises the tasks with the lowest rates. In effect, criticality
equals rate in RMPA.

The main alternative to RMPA is \emph{earliest deadline first} (EDF)
scheduling. This is a dynamic priority scheme, which at any time
schedules the task with the closest deadline. Unlike fixed-priority
schemes, such as RMPA, EDF is optimal on a uniprocessor in that it can schedule
any task set, as long as the total utilisation does not exceed
100\%. However, the dynamic prioritising implies that under overload,
EDF drops deadlines of all tasks \citep{Buttazzo_05}, meaning that there is no
concept of task criticality at all.

MCS require control over which deadlines will miss in the case of
overload: those of the tasks with low criticality (called \crit{low}
tasks from now on), while guaranteeing deadlines of \crit{high} tasks,
\emph{irrespective of scheduling priority}. This requires a mechanism for
limiting CPU time of high-priority tasks.

An established way of providing isolation is through \emph{scheduling
  reservations} \citep{Mercer_ST_93, Oikawa_Rajkumar_98}, where a
reservation \emph{guarantees} a certain share of the CPU to a periodic task. Such schemes are
popular in soft RT systems, e.g.\ multimedia, and some allow slack
time to be used by best-effort tasks \citep{Brandt_BLB_03}.
Scheduling reservations can be implemented as \emph{sporadic servers}
for RMPA \citep{Sprunt_SL_89} and with
\emph{constant bandwidth servers} (CBS) \citep{Abeni_Buttazzo_04}
on EDF.  \label{s:reservations}

Reservations present a guarantee by the kernel that the reserved
bandwidth is available. This means that they do not support
over-committing. Also, the kernel must perform a schedulability
analysis as \emph{admission control} whenever a reservation is
created. Schedulability tests can be complicated and frequently
constitute a trade-off between cost of the test and achievable
utilisation.

Recently the concept of a \emph{mode switch} was introduced to support
mixed criticality \citep{Burns_Davis_14}: when the system is unable to
meet its deadline, it enters a high-criticality mode, where the
priority of \crit{high} tasks is boosted above all \crit{low} tasks.
To achieve this, \crit{high} tasks are assigned multiple reservations, one per criticaly level.
In a two-criticality system, \crit{high} tasks have a pessimistic WCET
and an optimistic \emph{worst-observed execution time} (WOET). 
\crit{low} tasks have just one estimate. 
When the system is in \crit{low} mode, \crit{high} tasks run according to their WOET. 
If all reservations in this mode are schedulable, temporal isolation is guaranteed. 
However, if a \crit{high} task exceeds its WOET, the system switches to \crit{high} mode, degrading \crit{low} tasks and assuring assymetric protection between \crit{high} and \crit{low} threads without falsely correlating rate and urgency.

\subsection{Support for sharing and communication}\label{s:sharing}

As indicated in the introduction, integrity of critical components
must be assured even when tasks communicate and share. In our AAV
example of \autoref{f:aav}, the mission plan component encapsulates
waypoints. The \crit{high} flight-control component must be able to access a
consistent view of the flight plan, despite the \crit{lower} C\&C and other
components performing updates.

Encapsulating the shared data and the code that accesses and modifies
it into a single-threaded \emph{resource server}
\citep{Brandenburg_14} is a simple and effective way to achieve the
necessary transaction semantics. Obviously, this server has the criticality level of its most
critical client, but must also act on behalf of a \crit{low} client. This
creates a temporary \emph{criticality inversion} where the \crit{low} client
blocks the \crit{high} one. This is an unavoidable consequence of
sharing, and the design must ensure that it does not
cause the \crit{high} task to miss deadlines.

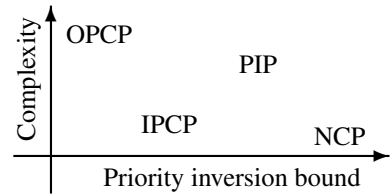
\begin{figure}[ht]
  \centering
  \setlength{\unitlength}{1mm}
  \begin{picture}(50,25)(-5,-5)
    \thicklines
    \put(-5,0){\vector(1,0){50}}
    \put(7,-4){Priority inversion bound}
    \put(0,-5){\vector(0,1){25}}
    \put(-4.5,2){\rotatebox{90}{Complexity}}
    \put(2,15){OPCP}
    \put(12,3){IPCP}
    \put(25,11){PIP}
    \put(35,1.5){NCP}
  \end{picture}
  \caption{Comparison of real-time locking protocols based on
    implementation complexity and priority inversion bound.}
  \label{f:locking}
\end{figure}

There are multiple ways to achieve mutual exclusion in fixed-priority
RT systems \citep{Sha_RL_90}, the most common being non-preemptive critical sections (NCP), the priority inheritance protocol (PIP), and the immediate\footnote{Also known as highest lockers protocol and \texttt{PRIO\_PROTECT} in POSIX.} and original priority ceiling protocols (IPCP and OPCP). For RT systems, the most important factor for mutual exclusion is the bound on priority inversion, where a low priority task blocks a high one. For efficient systems, the concern is execution cache performance, for secure systems
the concern is the avoidance of channels. 

\autoref{f:locking} shows the four protocols in terms of complexity
and priority inversion, none is a silver bullet.
NCP is simplest yet has the longest blocking time, IPCP requires the
priorities of all lockers to be known \emph{a priori}, PIP has high
implementation complexity and risks deadlock if resource ordering is
not used. OPCP is even more complex, and requires global state to be
maintained across all locks in the system, which is not acceptable for
seL4 as it introduces covert channels and is incompatible with seL4's
decentralised user-level resource management. We will show in
\autoref{s:sc} how IPCP can be easily implemented without the kernel
requiring knowledge about critical sections.

As a mechanism for supporting sharing,
Fiasco~\citep{Steinberg_04:dipl} introduced the idea of scheduling contexts, separate to 
execution contexts (threads). Scheduling contexts encapsulate priority, scheduling parameters and accounting detail,
and pass between threads over IPC. \citet{Steinberg_BK_10} extended this with bandwidth
inheritance~\citep{Lamastra_LA_01, Lipari_LA_04, Faggioli_LC_10} over IPC. This is 
equivalent to PIP combined with reservations.
When an IPC from client $B$ arrives at a server $S$, who is serving a client $A$ with an expired budget, $B$ budget is used to complete $A$'s request such that $B$ does not have to wait for $A$'s reservation to be replenished.
This kernel-implemented policy, also referred to as \emph{helping},
prevents the server from choosing alternatives which might be more
appropriate in a particular situation, such as aborting $A$'s request. 

The Fiasco design of scheduling contexts \citep{Lackorzynski_WVH_12} is tied to the traditional L4 model of
sending IPC messages directly to threads, a model which
has been abandoned in modern L4 kernels (including Fiasco and seL4) as it introduces covert
channels~\citep{Shapiro_03}. It is not supported on the later,
capability-based Fiasco.OC kernel.

\composite~\citep{Parmer_West_08} completely frees the kernel from any scheduling policy by providing
mechanisms for hierarchical user-level scheduling. It reduces overhead-related capacity loss
by configuration buffers shared between user-level and the kernel.
Some capacity loss remains as timer interrupts must be delivered down
the scheduling hierarchy. This approach does not suit seL4, as the
required reasoning about concurrent access (by kernel and user-level)
to those buffers would drastically
increase verification overhead \citep{Klein_AEMSKH_14}.
 
Unlike all L4 microkernels, \composite implements a migrating thread
model~\citep{Ford_Lepreau_94}. This implies that access to shared resources
does not block, thus avoiding priority inversion, although at the cost
of requiring all server code to be re-entrant, which is fairly
heavy-handed policy for a microkernel. 
Also, it only shifts the problem, as mutual exclusion is
still needed, including a way of limiting priority inversions. Given
the challenges of getting concurrent code right, it should be
minimised in high-assurance systems.

Linux introduced an implementation of the POSIX
\texttt{SCHED\_DEADLINE} in 3.14, which implements EDF with CBS for
temporal isolation. However RT tasks in Linux are higher priority than
all other tasks in the system, and cannot be over-committed (although
cgroups allow limiting the RT class to a certain share of the CPU).
Quest-V~\citep{Li_WCM_14} and PikeOS~\citep{Kaiser_Wagner_07} are both separation kernels for multicore systems that dedicate
cores to different criticalities.
AUTOBEST~\citep{Zuepke_BL_15} is another separation kernel where the authors demostrate
implementations of AUTOSAR and ARINC653 in separate partitions.

\subsection{seL4}\label{s:sel4}


seL4 is a high-performance OS microkernel with an unprecedented degree
of assurance: it features formal proofs of implementation correctness
down to the binary, proofs of spatial isolation properties
(enforcement of confidentiality, availability and integrity) and a
complete and sound analysis of worst-case execution times on ARMv6 processors
\citep{Klein_AEMSKH_14}. This assurance makes seL4 an appealing
candidate OS for critical systems.

seL4 is designed to be a general-purpose platform,
supporting a wide range of use cases. This is a reason why it has a
strong emphasis on performance, as many of the envisioned deployment
scenarios are performance-sensitive (e.g.\ mobile devices). Formal
verification is a strong motivator for generality: the cost of
assurance is best amortised if all use cases are supported by the
same, unmodified kernel \citep{Heiser_Elphinstone_16}. As the
maintainers commit to re-verify any changes to the mainline kernel,
they are only interested in changes that make the kernel more general,
not more specialised.

\subsubsection{seL4 overview}

In line with the microkernel minimality principle
\citep{Liedtke_95}, seL4 only provides a small number of policy-free
mechanisms. Specifically it provides for threads, represented as
\emph{thread control blocks} (TCBs), \emph{address spaces}, which are thin wrappers
around hardware page tables, and \emph{frame} objects, which represent physical
memory that can be used to populate address spaces by \emph{mapping}.
It further provides port-like \emph{endpoint} objects for synchronous
(rendezvous-style) communication and \emph{notification} objects,
which are essentially arrays of binary semaphores.

Like other security-oriented systems, seL4 uses capabilities
\citep{Dennis_VanHorn_66} for controlling access to all spatial
resources and providing complete mediation similar to KeyKOS
\citep{Bromberger_FFHLS_92} and EROS \citep{Shapiro_SF_99}. Besides
its assurance story, seL4's most characteristic aspect is its
isolation-oriented approach to memory management, which is made
policy-free by fully delegating it to user level.

Specifically, the kernel never allocates memory. After booting, seL4
hands all rights to any unused memory to the first user process in the
form of capabilities to \emph{Untyped} memory. The only operation
supported on Untyped is to \emph{retype} into some other object type
(TCB, page tables, frames etc), or to \emph{revoke} of an earlier retype. That way
user-level managers have full responsibility for any memory
management. For example, the initial process can partition Untyped
memory into several disjoint pools, and set up secondary resource
managers in each partition. The partitions are then totally isolated,
unless the initial process also provides access to some shared
resources (e.g.\ frames or endpoints) to support communication.

Like any kernel operation (other than the \code{yield()} syscall which
simply forfeits the remainder of the present time slice), IPC and
notifications are authorised by capabilities: a thread needs an
endpoint capability in order to send or receive messages, and a
notification capability for signalling or collecting notifications.

Similar to other L4 kernels, the kernel not only supports basic 
\code{send()} and \code{receive()} operations, but also two
versions of a send followed by a receive in one atomic syscall. First
there is the RPC-like \code{call()}, which is typically used by a
client invoking a server. When invoking \code{call()} on an
endpoint, the kernel creates a single-use \emph{reply capability},
which refers to a virtual, temporary \emph{reply endpoint}. The kernel
delivers the reply capability to the receiver listening on the endpoint, and
makes the sender wait on the reply endpoint. 

The second combined call is \code{reply\_receive()}, which sends a
message to the (implicitly supplied) reply endpoint and then makes the
invoker wait on a new request on the endpoint specified in the
syscall. Once used in the reply, the reply endpoint and capability are
removed.

\subsubsection{Scheduling}\label{s:sel4sched}

Management of time is comparatively under-developed in seL4. It  presently
implements the same simplistic scheduling model used in most L4
kernels for 20 years: priority-based round robin. The only
controllable parameters are a thread's priority and time slice. This
is not sufficient for supporting MCS, as indicated by the examples
given in the introduction.

On IPC, seL4 uses a \emph{direct process switch}
\citep{Liedtke_93}  where  possible, to avoid the cost of invoking the scheduler: the
IPC switches context from sender to receiver, but with the receiver
running on the sender's time slice, until it replies or is
preempted. When rescheduled after preemption, the server will execute
on its own time slice (and after replying to the client, the latter
may execute on the server's time slice). The IPC paths which do not
require scheduler invocation are implemented by separate,
highly-optimised \emph{fast-path} code.

This form of time-slice donation \citep{Steinberg_BK_10}
has been criticised as inappropriate for RT systems
\citep{Ruocco_08}, as time is not accounted
properly. Consequently, the Fiasco L4 kernel allows the sender to
specify whether donation is permitted.

However, even without time-slice donation, traditional L4 scheduling
is problematic. Consider a typical scenario of two clients, \(A\),
\(B\), invoking server \(S\).
Both clients have the same priority, which is lower than the server's,
and the same time slice length, so they ought to get equal amounts of
time. Assume client \(A\) requests long-running operations from \(S\),
while \(B\)'s requests are short. The server's time is not accounted against the
clients, and \(A\) gets a much higher share of the system than \(B\).
Furthermore, if the scheduler is invoked on each IPC, \(A\) and \(B\)
will alternate execution after each server invocation, making it very
difficult to reason about the progress of individual
tasks. Alternatively, if \(A\) continues executing after the invocation of \(S\)
returns, then \(A\) can effectively deny \(B\)'s service by invoking \(S\)
in a tight loop. 

In summary, the L4 model of managing time is unsatisfactory no matter how it is
implemented. The bandwidth-inheritance approach taken in some
kernels \citep{Steinberg_BK_10} is not a good solution either for
the reasons explained in \autoref{s:sharing}: on the one hand there is
the general issue of complexity and poor priority-inversion bound of
inheritance. On the other hand, inheritance offers no policy flexibility on managing overruns in servers.
Additionally, while Fiasco's implementation of bandwidth inheritance
allows for bounded priority inversion,
it violates temporal isolation: \(A\) is allowed to consume \(B\)'s
budget. 

\subsection{Summary}

We want a model that is simple enough to be suitable for seL4,
provides temporal isolation, and provides freedom in the
implementation of policies for dealing with isolation violations. At
the same time, it must continue to support all existing or anticipated
use cases of the kernel.

\section{Scheduling Model}\label{s:design}

We now present a scheduling model for seL4 which satisfies all
requirements for MCS stated in \autoref{s:intro}. It is based on a
small number of abstractions, namely
\begin{compactitem}
\item periodic threads with hard CPU bandwidth limits
\item scheduling contexts
\item timeout exceptions
\item notion of criticality in addition to priority \& explicit mode switches.
\end{compactitem}
Our model matches the approach known as \emph{static mixed
  criticality}  in scheduling theory \citep{Baruah_BD_11}, which
provides appropriate tools for analysis.

\newlength{\Unit}\setlength{\Unit}{1em}
\newcommand{\Bx}[1]{\raisebox{0.7ex}{\fbox{\rule{#1\Unit}{0pt}\rule{-0.7em}{0pt}}}}
\newcommand{\Tk}{\rule{\Unit}{0pt}\rule{-1pt}{0pt}\rule{1pt}{0.2ex}}
\newcommand{\Sp}{\rule{\Unit}{0pt}}
\newcommand{\Wp}{}              
\begin{figure*}[t]\centering
  \subfloat[Two periodic RT threads plus one best-effort thread running in slack time.]{
    \begin{tabular}{@{}rrrrrc@{}}
      \bf P& \bf T&\bf B& \bf U& \bf u& \bf Schedule \\
      3 &   5 & 1 & 0.2 & 0.2 & 	\Bx{1}\Tk\Tk\Tk\Sp\Bx{1}\Tk\Tk\Tk\Sp\Bx{1}\Tk\Tk \\
      2 & 10 & 5 & 0.5 & 0.5 & 	\Sp\Bx{4}\Sp\Bx{1}\Tk\Tk\Tk\Sp\Bx{2}\Wp\\
      1 & 20 &20 & 1.0 & 0.3 &	\Tk\Tk\Tk\Tk\Tk\Tk\Sp\Bx{3}\Tk\Tk\Tk\\
    \end{tabular}
  }
  \hspace{5ex}
  \subfloat[Three full-budget threads scheduled as in traditional L4.]{
    \begin{tabular}{@{}rrrrrc@{}}
      \bf P& \bf T&\bf B& \bf U& \bf u& \bf Schedule \\
      2 &  1 & 1 & 1.0 & 0.5 &	\Bx{1}\Sp\Bx{1}\Sp\Bx{1}\Sp\Bx{1}\Sp\\
      2 &  1 & 1 & 1.0 & 0.5 &	\Sp\Bx{1}\Sp\Bx{1}\Sp\Bx{1}\Sp\Bx{1}\\
      1 &  1 & 2 & 1.0 & 0.0 &	\Tk\Tk\Tk\Tk\Tk\Tk\Tk\Tk\\
    \end{tabular}
  }
  \caption{Examples of thread schedules. P=priority, T=period,
    B=budget, U=max.\ utilisation, u=actual utilisation.}
  \label{f:budget-ex}
\end{figure*}

\subsection{Execution-time limits: Budgets}

A key observation from \autoref{s:mc-sched} is that pure
priority-based scheduling cannot satisfy the requirements of MCS, and
we need a mechanism for temporal isolation. To achieve this we
introduce the notion of a \emph{budget}, which is a \emph{hard limit} on the
time a thread can consume during a \emph{period}. The ratio of budget
over period is the limit of CPU bandwidth a thread can consume.

Budgets are similar to the reservations introduced in
\autoref{s:reservations}, except that the kernel makes \emph{no guarantee
that any bandwidth is achieved}, only that the limit is not
exceeded. This makes admission control a user-level responsibility,
avoiding any policy in the kernel about whether admission should be
determined on- or off-line, should be static or dynamic, or should be
hierarchical of flattened \citep{Lackorzynski_WVH_12}. In particular,
the system designer may decide to trust a particular task not to use
its budget (except in emergencies) and perform the schedulability
analysis based on that knowledge.

\emph{Despite providing weaker guarantees, budgets are a more powerful
concept than reservations.} Specifically, if a set of reservations is
schedulable,  i.e.\ admission control succeeds, then budgets will
produce the same schedule, i.e.\ they behave like reservations. If,
however, the total is not schedulable, but the sum of all budgets
above some threshold priority \(p\) is, then all budgets of tasks whose
priority exceeds \(p\) still behave like reservations, but nothing of
priority \(\leq p\) is guaranteed any CPU time. 

This property allows us to safely overload a system with predictable
outcomes and without the kernel performing any admission control. We
will see later how this example of \emph{less is more} allows us to support MCS.

Specifically, we replace the kernel's notion of a \emph{time slice} by
two new attributes: \emph{period} and \emph{budget}. The budget is
less than or equal to the period and the ratio specifies the maximum
share (utilisation) of the CPU the thread can possibly get.  This is essentially the
model of sporadic servers introduced by \citet{Sprunt_SL_89}, except
that we use budgets instead of reservations.

The operation of the seL4
scheduler changes only slightly: it still picks the highest-priority
runnable thread, using round-robin within a priority. The difference
is that when the kernel schedules a thread, it sets a timer to enforce
the budget, and a thread whose budget is expired is no longer
runnable. The period specifies when the thread's budget is
replenished, thus making it runnable again. \autoref{f:budget-ex} shows
some examples.

Similarly to the budget not guaranteeing any time, the
period does not guarantee that a thread is actually scheduled
periodically (which depends on the priorities, periods, and budgets of
all other threads with the same or higher priority). Note also that
this model exactly emulates the existing seL4 scheduler when all
budgets are \emph{full}: if every thread has a budget that is equal to its
period, the period has the same semantics as the time slice used to
have.

\subsection{Scheduling contexts}\label{s:sc}

In order to provide better control over the time resource, we
introduce a \emph{scheduling context} (SC) object that grants access
to time. Instead of a time slice, a thread (in its TCB) holds a
scheduling context capability (scCap), without such a valid scCap, the thread is not
runnable.

The SC consists of the period, budget pair introduced above and thus
represents the maximum bandwidth a thread may consume.
The semantics of SCs are equivalent to 
hard reservations in Linux/RK~\citep{Rajkumar_JMO_98}, in that once the budget is exhausted,
no thread can run on that SC until it is replenished, however they differ in two ways.
First, we only allow one thread per SC at a time, but SCs can be passed between threads 
via IPC for cooperative scheduling. 
This allows for a minimal, single level scheduler in the kernel.
Second, the kernel does not conduct an admission test. Our SCs differ
from those of NOVA \citep{Steinberg_BK_10} in that priority remains
a thread attribute instead of being associated with an SC, and we
allow only one SC per thread.

SCs are like other seL4 objects, in that any thread that can allocate memory
can create them. However, setting the budget requires
special privilege, as creating budgets amounts to control over the
right to consume CPU time. It must be authorised by a capability.

We use an approach that is analogous to managing interrupt sources in
seL4. Specifically, there is a per-core virtual scheduling-control
object, represented by the \code{sched\_control} capability. This capability
must be presented when setting the budget of an SC. The
kernel creates this capability at boot time and hands it to the initial task
as part of the startup protocol.

\begin{figure}[htb]
  \centering
  \includegraphics[width=0.8\columnwidth]{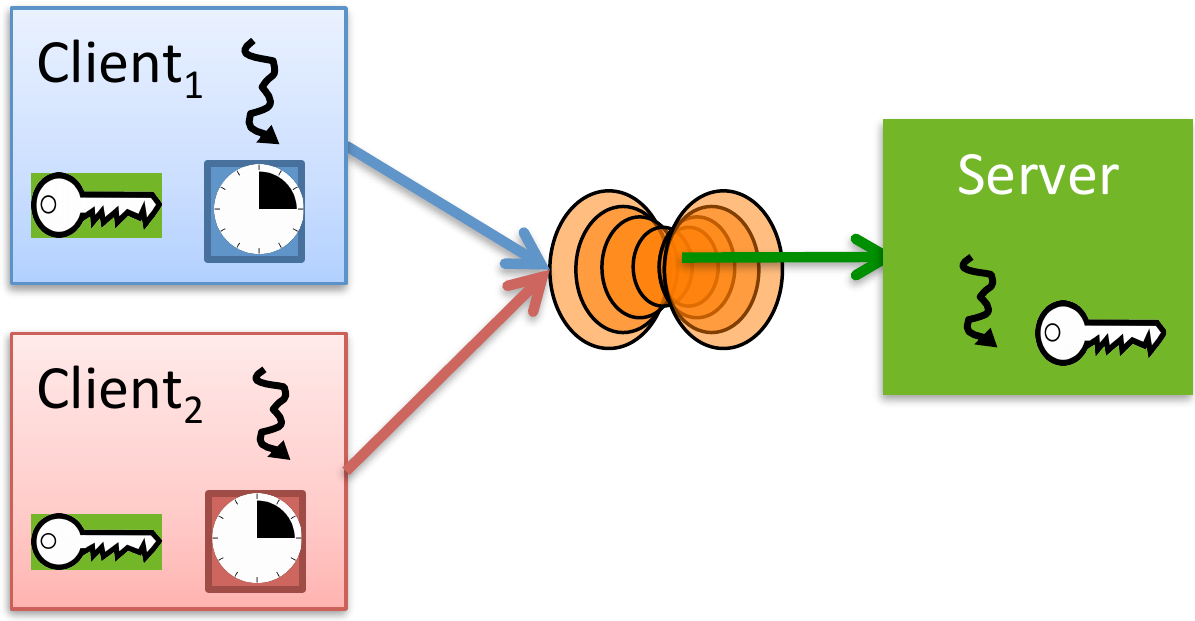}
  \caption{Resource server implementation as a ``passive'' thread
    without a scheduling context.}
  \label{f:passive}
\end{figure}

SCs provide a clean solution to the shared-server accounting dilemma
outlined in \autoref{s:sel4sched}. We allow a (server) thread without
an scCap  to wait on an endpoint, we call this a \emph{passive
  server}. If a client sends a message to this
endpoint, the IPC will transfer the client's scheduling context to the
server, which then executes on the client's \emph{borrowed} budget. The SC returns to
the client when the server replies to the client request. An example
is given in \autoref{f:passive}, where an SC-less resource server has
two clients, each holding an SC (indicated by the clock dial
representing a CPU bandwidth bound). 
For security, the sender must agree to the donation, the IPC will fail if
the receiver has no SC but the sender is unwilling to lend its own. No
SC transfer takes place if the receiver has its own SC (active
server).

A passive server can trivially implement the immediate priority
ceiling protocol introduced in \autoref{s:sharing}, by setting its
priority to the ceiling of priorities of all clients. As any client
needs a send capability on the server's endpoint, usermode managers
can control access to the server, and thus enforce the priority
ceiling.  We discuss in \autoref{s:timeout} how we deal with the server running
out of budget.

\begin{figure}[htb]\small
\hspace*{0.5em}%
\begin{minipage}{0.5\columnwidth}
\begin{verbatim}
notification_t ntfn;
sched_context s_sc;
tcb_t s_tcb;

void init() {
  // bind SC to TCB
  bind(s_sc, s_tcb);
  // create server thrd
  start_thread(s_tcb);
  // block and allow
  // server to run 
  wait(ntfn);
  // server initialised 
  // convert to passive
  unbind(s_sc);
}
\end{verbatim}
\end{minipage}
\begin{minipage}{0.4\columnwidth}
\begin{verbatim}
notification_t ntfn;
endpoint_t ep;

void server() {
  // run on init SC
  initialise();
  // signal & block
  signal_recv(ntfn, ep);
  while (true) {
    // run on client SC 
    process_request();
    // reply & block
    reply_recv(ep);
  }
}
\end{verbatim}
\end{minipage}
\caption{Passive server initialisation.}
  \label{f:init}
\end{figure}

Passive servers must be initialised with an initialisation SC
and then communicate to the initial task when they are done, 
such that the SC can be removed. We support this with
a new system call \texttt{signal\_receive()}, which combines
signalling a notification with an IPC receive. \autoref{f:init} shows
how initialisation works in principle.

Call-reply\&wait IPC with SC transfer avoids invoking the scheduler or
updating accounting data during IPC, and thus retains the low overhead
of the direct process switch optimisation. It has in fact many of the properties of a
migrating thread model \citep{Ford_Lepreau_94}, specifically it
avoids having multiple schedulable entities for what is logically a
single-threaded operation. The advantage over migrating threads is
that the kernel does not have to provide stacks on the fly, and thus is free
of policy decisions such as determining stack sizes, charging for
memory, whether to cache stacks. Instead, our model requires explicit
user-level management of stacks through thread objects.

\FIXME{Mention that scheduling theory just works, with budget=WCET.}

\subsection{Managing thread execution}

A periodic thread needs to suspend itself when it has finished
processing for the current period. It does so by calling
\code{yield()} on its own SC.\footnote{Authorising \code{yield()} with
  an SC capability removes seL4's
previous anomaly of having a syscall that requires no capability to
execute. \code{yield()} can also be called on another thread's SC,
  cancelling that thread's current budget. However, we do not claim
  that there is a good use case for this.} An event-triggered (sporadic)
thread instead waits on its IRQ notification. Of course, even if the notification
is signalled (by an IRQ or another thread), the sporadic thread will
only execute if it has budget.

Sometimes explicit changes of a thread's priority are needed, e.g.\
when implementing IPCP without encapsulating the critical section into
a separate server. In order to change a thread \(B\)'s priority,
thread \(A\) must hold a capability to \(B\)'s TCB. In order to
prevent arbitrary priority changes, we re-introduce the concept of a
\emph{maximum controlled priority} (MCP) that was used in early L4
versions \citep{Liedtke_96:rm}. 
Specifically, \(A\) cannot \emph{raise}
any thread's priority, including its own, higher than \(A\)'s
MCP. Note that this does not stop \(A\) from having a priority higher
than its MCP, but some other thread must have set it up.

We add two further operations on scCaps in order to allow fine-tuning
scheduling decisions. The first, \code{consume}, obtains the total
time accounted the designated scheduling context since the last such
enquiry.

The second, \code{yieldto()}, allows user level to manipulate
the kernel's scheduling queues. When invoked on a scCap, \textbf{and} the
designated SC is presently associated with a thread whose priority
does not exceed the callers MCP, \textbf{and} the thread has budget
available in its present period, \textbf{then} that thread is moved to the head
of the ready queue of its priority. This ensures that it is the next
thread to be scheduled if no higher-priority threads are
runnable. Invoking \code{yieldto()} implicitly invokes \code{consume},
i.e. it returns and resets the time accumulated on the SC.

\subsection{Budget overrun}\label{s:timeout}

We provide \emph{timeout exceptions} in order to detect budget
overrun, analogous to seL4's treatment of other exceptions. An
seL4 thread already has an exception 
endpoint. If an exception is triggered, the kernel sends a message to
the appropriate endpoint on the faulting thread's behalf. An exception
handler waiting on the endpoint will then receive the message and take
appropriate action. By replying to the exception message, it unblocks
the faulting thread (possibly after adjusting its instruction pointer
to skip an emulated instruction). In practice, many threads share the
same exception endpoint (and thus handler).

We extend this model by adding a timeout-exception endpoint: the
kernel sends a message to that endpoint when the thread exceeds its
budget, and the handler can take appropriate action, which may include
adjusting the faulting thread's budget. If the handler increases the
budget and then replies to the fault message, the thread will continue
to run on the remainder of the enlarged budget.
A thread without a timeout-exception endpoint is simply rate limited.

Timeout exceptions allow recovering from priority/criticality
inversions, as possible in the passive resource server of
\autoref{f:passive}. If the server's borrowed SC runs out of budget,
its timeout handler can implement appropriate policy, such as letting
the server complete the request on an emergency budget, forcing a
reset or roll-back, possibly
coupled with taking some additional safety precautions prejudicial to
C\(_1\), such as suspending C\(_1\) or affecting a criticality mode
switch. This is in contrast to kernel-implemented helping schemes,
which implement a specific policy.

\begin{table}[th]\centering
  \begin{tabular}{|lrrrrr|}
    \hline
    	  	   &\bf C & \bf P& \bf T&\bf B     & \bf U\\
    \hline
    \rowcolor{Lavender}
       \(T_5\) &       1 &      6 &    10 &     2      & 0.20 \\
    \rowcolor{Lavender}
       \(T_4\) &       1 &      5 &    20 & \(2|7\) & \(0.1|0.35\) \\
    \rowcolor{SkyBlue}
       \(T_3\) &       0 &      4 &    25 &     5      & 0.20 \\
    \rowcolor{Lavender}
       \(T_2\) &       1 &      3 &    40 &     4      & 0.20 \\
    \rowcolor{SkyBlue}
       \(T_1\) &       0 &      2 &    60 &     6      & 0.20 \\
       \(T_0\) &       0 &      1 &  100 &  100     & 0.00 \\
    \hline
  \end{tabular}
  \caption{Parameters of a sample system, where \(T_4\) has a
    \crit{low} budget of 2 and a \crit{high} budget of 7. C=criticality, P=priority, T=period,
    B=budget, U=utilisation.}
  \label{t:mc-params}
\end{table}

\subsection{Criticality mode switches}

Another case of budget overrun is the system shown in \autoref{t:mc-params},
consisting of three \crit{high} tasks (pink) and two
\crit{low} tasks (blue), plus \(T_0\) which runs in slack time. With
\(T_4\)'s \crit{low} budget of 2~units, this system is RMPA
schedulable --- the RMPA utilisation bound for 5 tasks is 74\% --- so all tasks
will meet their deadlines.

Now assume that \(T_4\) overruns its budget, triggering a timeout
exception. The handler can adjust its budget to the \crit{high} value
of 7~units, however, the resulting system is no longer
schedulable. Since the 4-task utilisation bound of RMPA is 75\%, not
only the \crit{low} task \(T_1\) may miss its deadlines, but also the
\crit{high} task \(T_2\).

We can repair this situation by a criticality switch that prevents \crit{low} tasks,
specifically \(T_3\) from competing with \(T_4\). 
We support this by introducing an explicit
notion of a \emph{system criticality level}, as well as a new
\emph{thread criticality} attribute. When setting the criticality
system level to \(C\), we boost the priority of all threads with
criticality \(\geq C\) by a constant amount, so that they all have
priorities above any lower-criticality threads \citep{Burns_Baruah_13}. In the above example,
the timeout handler not only increases \(T4\)'s budget, but also
raises the criticality level to one.

The lowest-priority task \(T_0\) will only run if there is slack in
the system. If so, the criticality level can be reset to zero
(possibly after waiting for a few of \(T_0\)'s periods).

We control thread criticality changes similarly to priority changes: a
thread attribute \emph{maximum controlled criticality} (MCC)
determines limits how a thread can chance another thread's
criticality, just as the MCP limits priority changes. Setting the
system criticality level requires the \code{sched\_control} capability.

\section{Implementation}\label{s:impl}

\subsection{Objects and methods}

We add a new 64-byte scheduling context object type, and modify global
state by eight words plus the number of criticalities.
In TCB objects we replace the \texttt{timeslice}) by the scCap, add a
timeout handler capability,
criticality, MCP, and a number of bookkeeping fields, a total of nine
extra fields. As TCB objects must be powers of two in size, this has
no effect on the size of a TCB object.


We add three methods on TCBs. SCs have 5 methods, the new
\code{sched\_control} has two. There are also three new methods for manipulating
reply capabilities: the ability to set your reply slot, save another threads reply capability, 
and the ability to swap your reply capability with one saved earlier. 
This extra flexibilty with reply capabilities allows for more efficient
user-level scheduling via IPC, and allows the 
timeout handler to access the faulter's reply capability, so it can
unblock the client on the server's behalf. Also new is \code{nbsend\_wait}.

\subsection{Scheduling algorithm}\label{s:sched}

Baseline seL4 has a ready queue, which satisfies the invariant that it
contains all runnable threads except the one presently executing
\citep{Blackham_SH_12}. It is implemented as a priority-indexed array
of queues. A two-level bitfield of occupied priorities ensures O(1)
access. 

The main change required to the existing seL4 scheduler is the
addition of a \emph{release queue}. A thread whose budget expired
before its period is up is removed from the ready queue and inserted
into the release queue. This retains the existing invariant for the
ready queue, while the release queue is characterised as holding all
threads that would be runnable but are presently lacking budget. The
queue is ordered by the time of the threads' next budget refresh,
i.e. the time their next period is up.

Whenever the kernel schedules a thread, it sets the timer to fire when
the thread's SC's remaining budget is due to expire, or for the next
wake-up time for the head of the release priority queue (whichever is first). If an SC switch
occurs, because the timer fires or the thread blocks without an SC
transfer, the consumed time is subtracted from the SC's budget and
added to the accumulated time. 

On kernel entry (except on the fastpath, which never leads to an SC
change or scheduler invocation) the kernel updates the current
timestamp and stores the time since the last entry. It then checks
whether the thread has sufficient budget to complete the kernel
operation. If not, the kernel pretends the timer has already fired,
resets the budget and adds the thread to the release queue. 

This adds a new
invariant that any thread in the scheduling queues must have enough budget to exit the kernel.
This makes the scheduler precision equal to the kernel's WCET, which for
seL4 is known (unlike any other protected-mode OS we are aware of).

Threads are only charged if the scheduling context changes, in order to avoid
reprogramming the timer which can be expensive on many platforms. 
Else, the timestamp update is rolled back by subtracting the 
stored consumed value from the timestamp.

\subsection{Criticality}

\newcommand{\Ncr}{N_\mathit{crit}}

A core integrity requirement of MCS is that the timeliness of
\crit{high} tasks is unaffected by low tasks. This includes the mode
switch: its cost must not depend on the number of \crit{low} tasks in
the system. We implement criticality as follows.

The kernel supports base priorities in the range \([0,2^{N_p}-1]\),
where \(N_p\) is a kernel build option. The base priority is the thread's actual
priority at system criticality level zero.
The number of criticality levels, \(\Ncr\), is also a build option. Typically,
it is a small number, e.g. \ \citet{DO178B} specifies five
levels. We require that \(\Ncr \times 2^{N_p} \leq 1024\).

For each criticality level the kernel maintains a queue of threads, threads
that are explicitly suspended (as opposed to out of budget or blocked
in IPC) are not in any criticality queue.

When system criticality changes from $C$ to $C'$, the kernel iterates through
the criticality queues from $C'$ to $\Ncr-1$. For each thread in those
queues, the kernel changes the present priority \(P\) to $P_0 \land
(C' \ll 8)$. This ensures that the
priority of all \crit{high} threads is above those of all \crit{low}
ones (with respect to $C'$).

The per-priority ready queues are doubly-linked lists of TCBs, so
moving a thread from one queue to another is a constant-time
operation. Hence, the total time for the priority adjustments is
proportional to the number of threads at criticality $C'$ or higher.

If during a criticality increase the kernel detects any threads that are running on a borrowed
scheduling context (comparing \texttt{tcb->sc->home} to \texttt{tcb}),
and the SC's owner is \crit{low} (\texttt{tcb->sc->home->crit} $\leq
C'$), it generates timeout exception for that thread. This allows a
server to abort any operation on behalf of a \crit{low} thread. If the
thread running on an SC borrowed from a \crit{low} thread has no
timeout hander, it will complete normally. In this case, the
worst-case blocking time is the worst-case server request time, plus
the cost of the mode switch.



\begin{table}[h!]\centering
  \begin{tabular}{|l|l|l|}\hline
                   & Co-operative & Preemptive           \\\hline
 Shared SC         & IPC          & Timer notifications  \\\hline
 SC per TCB        & Signals      & Timeout execeptions \\\hline

  \end{tabular}
  \caption{Mechanisms for user-level scheduling}
  \label{t:ul-sched-mech}
\end{table}

\subsection{User-level scheduling}

The kernel provides fixed-priority scheduling with budgets. This is a
particular (although quite flexible) policy. Fortunately, our
mechanisms allow us to implement very general policies, as indicated
in \autoref{t:ul-sched-mech}. 

For example, cooperative scheduling with an arbitrary policy can be
implemented with a shared SC, where the threads cooperate via IPC, or
per-thread SCs, where synchronisation is via notifications (although
it is unclear why one would want the latter). Pseudocode for both
variants is shown in \autoref{f:coop}.

Similarly, arbitrary preemptive scheduling policies can be
implemented, \autoref{f:preempt} shows pseudocode for schedulers with
shared or per-thread SCs. The shared-SC case uses one SC for all
threads, and a separate one for the timer.

\begin{figure}[htb]\small
\hspace*{0.5em}%
\begin{minipage}{0.5\columnwidth}
\begin{verbatim}
void coop_sched_s() {
    reply_recv(ep);
    p = t;
    t = pick_thread(p);
    swap_caller(t, p);
}

void coop_yield_s() {
    //yield
    call(ep);
}
\end{verbatim}
\end{minipage}
\begin{minipage}{0.4\columnwidth}
\begin{verbatim}
void coop_sched_m() {
    t = pick_thread(t);
    signal(t->ntfn);
    yieldTo(t->sc);
}


void coop_yield_m() {
    //yield
    wait(ntfn);
}
\end{verbatim}
\end{minipage}
\caption{User-level cooperative scheduler and thread yield function
  using a shared SC (left) and per-thread SCs (right).}
  \label{f:coop}
\end{figure}

\begin{figure}[thb]\small
\hspace*{0.5em}%
\begin{minipage}{0.5\columnwidth}
\begin{verbatim}

void pr_schd_s(prev) {
    // wait for timer 
    wait(timer);
    t = pick_thread();
    // change sc over
    swap_sc(t, prev);
    program_timer();
    ack_irq();
}
\end{verbatim}
\end{minipage}
\begin{minipage}{0.4\columnwidth}
\begin{verbatim}

void pr_schd_m() {
    // wait for timeout
    recv(ep);
    t = pick_thread();
    // place at head
    // of prio queue
    yield_to(t);
}

\end{verbatim}
\end{minipage}
\caption{User-level preemptive scheduler with shared (left) and
  per-thread (right) SCs.}
  \label{f:preempt}
\end{figure}

  \FIXME{At some point we need to argue the case for keeping a scheduler in the kernel at all - COMPOSITE doesnt have one at all}

\section{Evaluation}\label{s:eval}

We conducted our evaluation on two machines, both configured to use one core:
\begin{compactitem}
\item \textbf{Sabre:} 1\,GHz ARM Cortex~A9 system on chip on
  a Freescale i.MX6 SABRE Lite development board.
\item \textbf{Haswell:} 3.1\,GHz Haswell E1220v3 processor in a server
  machine running in 32-bit mode (64-bit seL4 is in development).
\end{compactitem}

\subsection{Microbenchmarks}

\begin{figure}[t]\centering
\begin{tabular}{|c|l|l|l|l|}\hline
\textbf{Arch}                 & \textbf{Operation}       &  \textbf{Baseline} & \textbf{RT} & \textbf{Diff} \\ \hline
\multirow{5}{*}{Sabre}        
             & \texttt{call()}      &  279             & 282       & +1\% \\ \cline{2-5}  
             & \texttt{replyrecv()} &  291             & 311       & +7\% \\ \cline{2-5}  
             & IRQ latency          &  467             & 578       & +24\% \\\cline{2-5}
             & \texttt{signal()}    &  107             & 111       & +4\% \\\cline{2-5}
             & Schedule             &  875             & 1242      & +42\% \\\hline
\multirow{5}{*}{Haswell}        
             & \texttt{call()}      &  412             & 415       & +1\% \\ \cline{2-5}  
             & \texttt{replyrecv()} &  414             & 426       & +3\% \\ \cline{2-5}  
             & IRQ latency          &  952             & 1448      & +44\% \\\cline{2-5}
             & \texttt{signal()}    &  383             & 387       & +1\% \\\cline{2-5}
             & Schedule             &  972             & 1532      & +58\% \\\hline

\end{tabular}
\caption{Microbenchmarks of seL4 baseline vs. RT kernels, standard
  deviations are negligible.}
\label{t:micro}
\end{figure}

\subsubsection{Kernel microbenchmarks}

\autoref{t:micro} shows the cost of the (performance-wise) most
important kernel operations of our present implementation compare to
the baseline seL4 kernel. Latency of the main IPC send+receive
operations increases by three cycles (call) and by 12--20 cycles
(reply\&wait). These are the result of extra checks on the fastpath to
accommodate scheduling contexts and ordering IPC, but the increase in cost is clearly
negligible. The same can be said for signalling a notification.

The actual cost of the model can be seen in the IRQ and scheduler
latency. Part of that is due to the need to reprogram the timer to
enforce the budget, which is needed on every scheduler invocation, but
also on an IRQ, as this normally unblocks a waiting handler. We
measure the cost of reprogramming the timer to be 55~cycles on the
Sabre, but about 200~cycles on the Haswell.

The rest of the increase is the result of the significant extra code
from dealing with scheduling contexts. Note that scheduling is
considered an expensive operation in seL4, and happens much less
frequently than IPC.

\subsubsection{Mode switch}

\begin{table}[h]
\centering 
\begin{tabular}{|c|c|r|r|r|r|}\hline
\textbf{Criti-} &\textbf{Threads} & \multicolumn{2}{c|}{\textbf{ARM}} & \multicolumn{2}{c|}{\textbf{x86}} \\
\textbf{cality} &\textbf{boosted} & \textbf{up} & \textbf{down} & \textbf{up} & \textbf{down} \\\hline
3 &    4 & 1.4$\mu$s & 1.7$\mu$s & 0.4$\mu$s & 0.5$\mu$s \\\hline
2 &    12 & 2.4$\mu$s & 2.4$\mu$s & 0.5$\mu$s & 0.6$\mu$s \\\hline
1 &    28 & 4.3$\mu$s & 3.7$\mu$s & 0.8$\mu$s & 0.7$\mu$s \\\hline
\end{tabular}
\caption{Results of switching from criticality level 0 to the
  criticality listed in column 1. Column 2 shows the number of tasks
  that need boosting. Standard deviations are no more than 2\%.}
\label{t:mode-switch}
\end{table}

To evaluate the cost of changing the system criticality level,
configure the kernel with 256 priorities and 4 criticality levels
(0--3). We then set up a system with 60 threads, of which 32, 16, 8
and 4 have criticality 0, 1, 2 and 3
respectively. 

\autoref{t:mode-switch} shows the cost of switching
criticality level between zero and one of the other levels. For each
data point we took 10,000 measurements with a primed cache.

\autoref{t:mode-switch} shows the cost of switching criticality level
between zero and one of the other levels. As the table shows, when
switching to level three, the three threads at that level need to be
boosted, while a switch to level one requires boosting all 28 threads
of criticality greater than zero.

The results show that a mode switch is fairly fast, around
1,500~cycles on both platforms as long as the affected number of
threads is small (which is to be assumed for \crit{high} threads), and
cost is roughly linear in the number of threads to be boosted.  This
is important, as the schedulability analysis must allow for that
cost. However, the numbers shown in \autoref{t:mode-switch} are
hot-cache (best-case) numbers, while the criticality analysis must be
based on WCET.

\autoref{t:mode-switch} shows the number of threads for each criticality and the results of the microbenchmark.

\subsection{Case studies}

\subsubsection{Linux CFS}

As an example of a complex dynamic-priority scheduling policy
implemented at user level, we implement a version of Linux' so-called
\emph{completely fair scheduler} (CFS). The implementation uses a
red-black tree and calls to \code{consumed()} to adjust the
weights. The scheduler runs one seL4 priority above its clients.

\begin{figure}[t]
  \centering
  \includegraphics{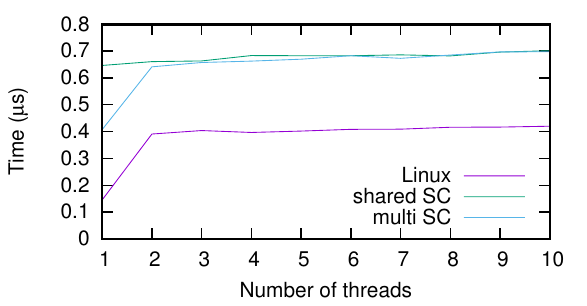}
  \caption{Execution time of yield operation measured from user-level
    for CFS compared with Linux yield on Haswell. Standard deviations
    are 2--3\%.}
  \label{f:cfs}
\end{figure}

\autoref{f:cfs} shows the scheduling cost for two scenarios, shared
and per-client SDC. Cost is measured by taking a
time stamp in the client, which then calls \code{yield()}, with
another time stamp taken right after (in the next client thread). 

We also show the cost of the same operations under Linux, which takes
about 50--60\% of the time. However, this turns out to be mostly the
syscall cost, as the Linux \code{yield()} bypasses the scheduler. So,
our user-level implementation looks quite competitive.

\subsubsection{EDF scheduler}

\begin{figure}[t]
  \centering
  \includegraphics{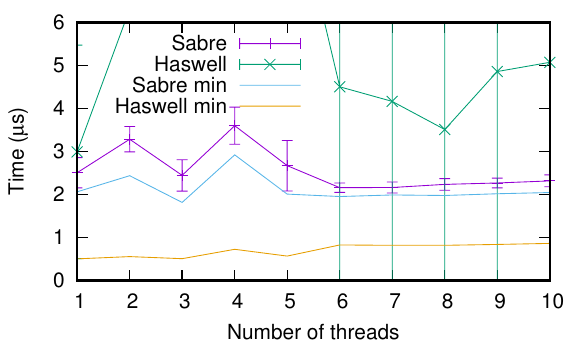}
  \caption{Execution time of EDF on Sabre and Haswell.}
  \label{f:edf}
\end{figure}

As a second scheduling policy we implement EDF at user level, this
time only the scenario with a shared SC. The results are shown in
\autoref{f:edf}. The standard deviations are very big, especially on
the Haswell platform. This is not unexpected, as the amount of work
EDF has to do on each scheduling operation is very sensitive to the
present state of the deadline and release queues.
The scheduler may have to release threads, reprogram the timer for the next release,
ack the previous interrupt and IPC the next thread, or resume a preempted thread with 
\texttt{yieldto}.

We used the \emph{randfixedsum}~\citep{Emberson_SD_10} algorithm to generate 10 EDF
task sets for each of the 10 data points, with periods between 10--100ms. Each task 
set ran 1000 times, for 10,000 runs for each data point.

A better metric in this case is the minimum scheduler time, shown in
the figure as ``Sabre min'', ``Haswell min''. It is reasonably stable
around 2\,\(\mu\)s for the Sabre, and 0.5--0.9\,\(\mu\)s for the
Haswell platform. This is an excellent result:
\citet{Cerqueria_Brandenburg_13} measured the latencies of various
in-kernel Linux schedulers on a Xeon~X7550 platform and found the
minimum to be around 1.5\,\(\mu\)s for all schedulers. While
comparions across different hardware must be taken with a grain of
salt, the fact that latencies of our user-level implementation is a factor four
less indicates that our performance is competitive.

\begin{figure}[t]
  \centering
  \includegraphics{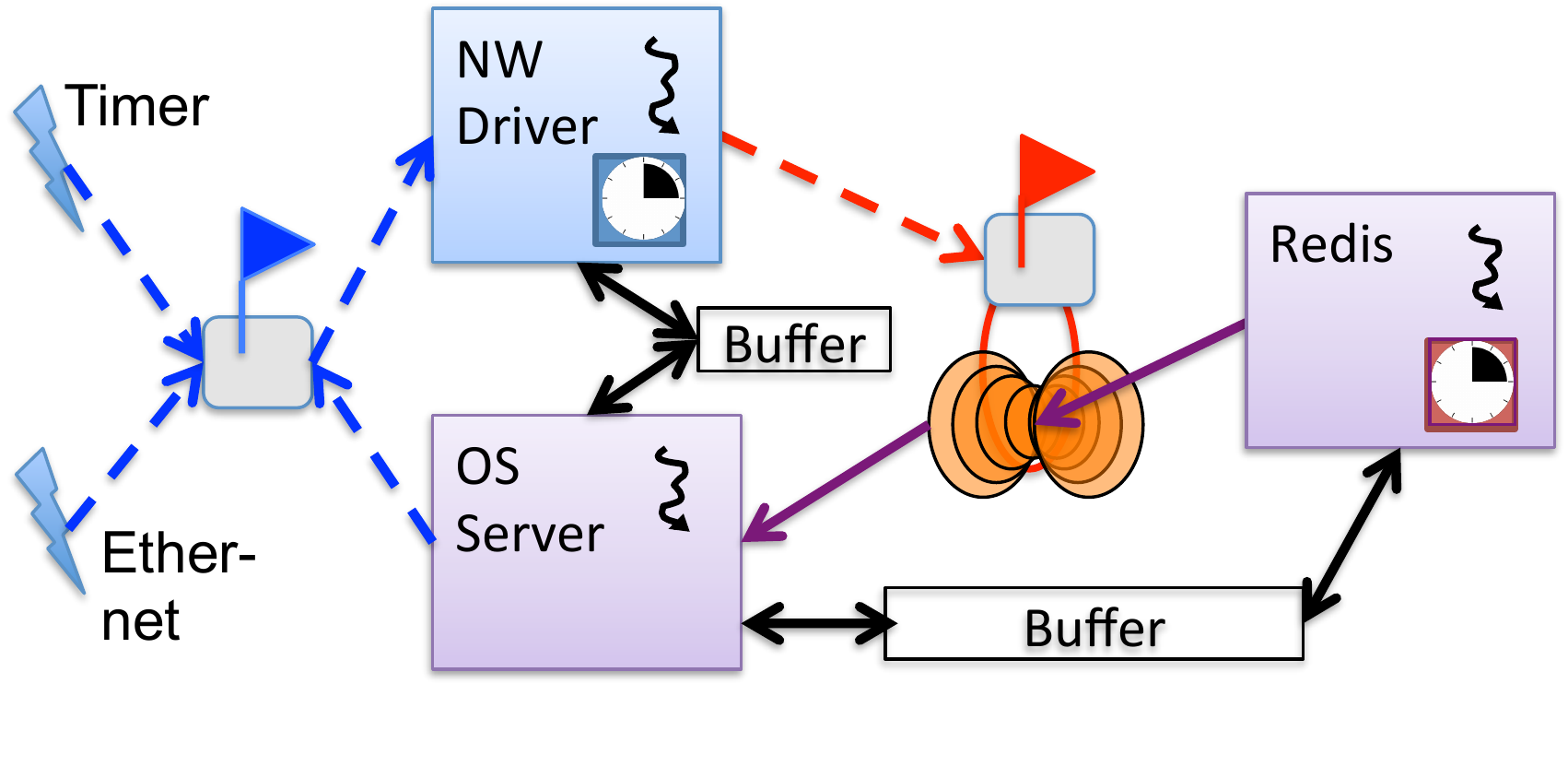}
  \caption{Network server setup, explanation in text.}
  \label{f:redis-arch}
\end{figure}

\subsubsection{Network server}

In order to demonstrate temporal isolation, we use a network
benchmark, specifically the Yahoo! Cloud Serving Benchmarks (YCSB)
\citep{Cooper_STRS_10}. We run this against a server using the Redis
key-value store \citep{redis:url}.

The server setup is shown in \autoref{f:redis-arch}. Dashed arrows
show synchronisation operations through notifications (semaphores)
indicated by flags, with coloured, broken single arrows indicating the
direction of the signal. The OS server, which contains the IP stack and is
implemented as a passive server, presents a POSIX interface, which is
implemented by an RPC protocol through an endpoint. The (active) Redis
server invokes the OS server (coloured, solid single arrow), which
then runs on Redis' scheduling context. Redis and the OS share a
buffer for passing bulk data (black, solid double arrows). The OS also
shares a buffer with the Ethernet driver, which uses a second
notification (red) for signalling completion to the OS. That
notification is ``bound'', meaning the signals are delivered to the
waiting OS as an IPC apparently coming from the endpoint.

\FIXME{12\% overhead between RT kernel and baseline}

    \begin{table}[th]
      \centering
      \begin{tabular}{|l| c | c | c |}
        \hline
        \bf Thread& \bf Prio & \bf Period & \bf Budget \\
        \hline
        Hog &	254 & 1\,ms & variable \\
        Driver & 253 & 2\,ms & 2\,ms \\
        Redis & 252 & 1\,s & 1\,s \\
        OS & 252 & -\,- & -\,- \\
        \hline
      \end{tabular}
      \caption{Scheduling parameters of network server setup.}
      \label{t:redis-param}
    \end{table}

Not shown is a
separate CPU hog thread, which does not communicate with this setup,
but is competing for CPU time. The hog runs at highes priority (254)
with a 1\,ms period. The Ethernet driver runs at priority 253

\begin{figure}[htb]
  \centering
  \includegraphics{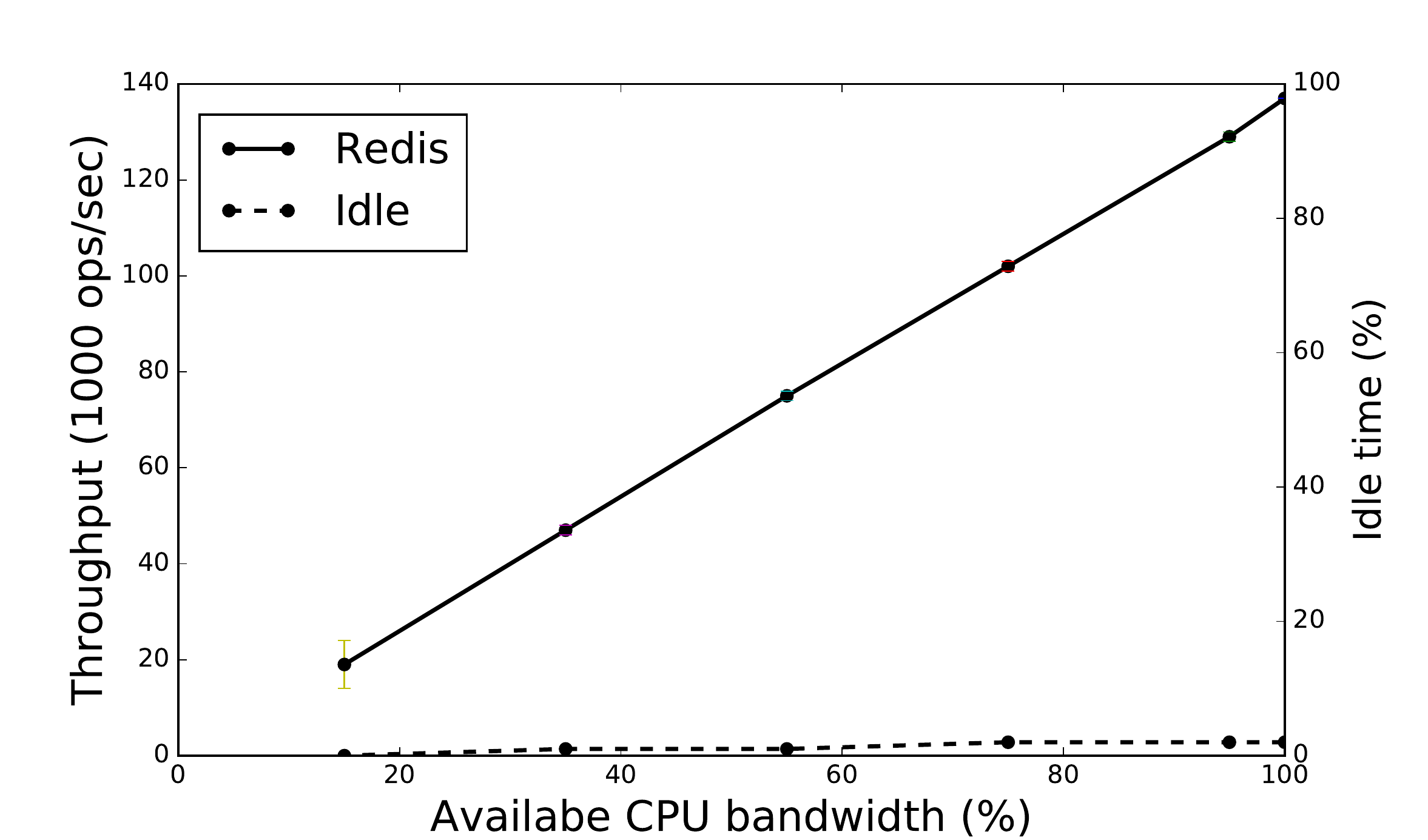}
  \caption{Throughput of Redis YCSB workload A vs available bandwidth,
    also showing idle time.}
  \label{f:redis}
\end{figure}

We use the budget of the hog to control the amount of time left over
for the server configuration. \autoref{f:redis} shows the bandwidth
achieved by the YCSB-A work load as a function of the available CPU
bandwidth (i.e.\ the complement of the bandwidth granted to the hog
thread). The figure also shows the total CPU idle time.

The graph shows that the server is CPU limited (very low idle time)
and consequently throughput scales linearly with available CPU
bandwidth.

\subsubsection{Server rollback}

As an example of a shared server running out of budget, we implement
the scenario of \autoref{f:passive} of a passive server with two
clients. The server is providing an encryption service using AES-256
using a block size of 16~bytes. The server alternates between two
buffers, of which one always contains consistent state, the other is
dirty during processing.

When the server runs out of budget, its timeout fault hander gets
invoked. It rolls the server back to the last consistent state and
makes it ready for the next client.

We measure rollback time, from the time the fault handler is
invoked, until the server is ready for the next request. Given the
small amount of rollback state, this measures the baseline overhead,
for servers with more state, the handling that state would
have to be added.

We run this on the Sabre and find a mean rollback time of 12 $mu$s,
with a 31\% standard deviation on 12 runs with a cold cache. The
individual times fluctuated between 9 and 24\,$mu$s.

\section{Conclusions and Future Work}\label{s:concl}

Mixed criticality systems are gaining traction in avionics and the
automotive sector, due to the SWaP issues created by mushrooming
functionality. In order to get the full benefit of MCS, we need an OS
supporting strong, but asymmetric temporal isolation. Inherent in the
notion of MCS is also a requirement for high assurance.

While there is a wealth of theory about MCS, little of it is
implemented in more than a proof-of-concept, certainly not in a
high-assurance OS. We have identified a model for temporal resource
management that lends itself to efficient and policy-free
implementation in a high-performance and high-assurance OS. We have
implemented this in seL4, and have demonstrated that the base model
supports the efficient implementation of a range of different
scheduling policies, and allows efficient handling of various
emergencies. 

Presently the main limitation of the work is the restriction to a
single core, which is the target of future work, as is the formal
verification of the real-time seL4 kernel.


\ifAnon\else
\section*{Acknowledgements}
The authors gratefully acknowledge Hesham Almatary's help in producing
the web-server benchmarks, and Bj\"{o}rn B. Brandenburg and Leonid
Ryzhyk for providing feedback on an early draft.  \fi

\balance
{ \sloppy
  \label{p:last}
  \bibliographystyle{plainnat}
  \bibliography{arxiv}
}
\end{document}